\documentclass[aps,pre,twocolumn,superscriptaddress,citesort]{revtex4-1} 

\usepackage{color}
\usepackage{graphicx}
\usepackage{CJK} 
\usepackage{amsmath}
\usepackage{epstopdf}
\usepackage[normalem]{ulem}
\begin{document}

	\begin{CJK*}{UTF8}{} 
		
		\title{
		Robustness of heat-transfer in confined inclined convection at high-Prandtl-number}
		

		\author{Linfeng Jiang}
		\affiliation{Center for Combustion Energy, Key Laboratory for Thermal Science and Power Engineering of Ministry of Education, Department of Energy and Power Engineering, Tsinghua University, Beijing, China}
		
		\author{Chao Sun}
		\email{chaosun@tsinghua.edu.cn}
		\affiliation{Center for Combustion Energy, Key Laboratory for Thermal Science and Power Engineering of Ministry of Education, Department of Energy and Power Engineering, Tsinghua University, Beijing, China}
		
		\author{Enrico Calzavarini}
		\email{enrico.calzavarini@polytech-lille.fr}
		\affiliation{Universit$\acute{e}$ de Lille, Unit$\acute{e}$ de M$\acute{e}$canique de Lille, UML EA 7512, F 59000 Lille, France}


		\date{\today}
		
		\begin{abstract}
We investigate the dependency of the magnitude of heat transfer in a convection cell as a function of its inclination by means of experiments and simulations. The study is performed with a working fluid of large Prandtl number, $\rm Pr\simeq480$, and at Rayleigh numbers $\rm Ra\simeq10^8$ and $\rm Ra\simeq5\times10^8$ in a quasi-two-dimensional rectangular cell with unit aspect ratio. By changing the inclination angle ($\beta$) of the convection cell, the character of the flow can be changed from moderately turbulent, for $\beta=0^o$, to laminar and steady at $\beta=90^o$. The global heat transfer is found to be insensitive to the drastic reduction of turbulent intensity, with maximal relative variations of the order of $20\%$ at $\rm Ra\simeq10^8$ and $10\%$ at $\rm Ra\simeq5\times10^8$, while the Reynolds number, based on the global root-mean-square velocity, is strongly affected with a decay of more than 85$\%$ occurring in the laminar regime. 	We show that the intensity of the heat flux in the turbulent regime can be only weakly enhanced by establishing a large scale circulation flow by means of small inclinations. On the other hand, in the laminar regime the heat is transported solely by a slow large scale circulation flow which exhibits large correlations between the velocity and temperature fields. For inclination angles close to the transition regime in-between the turbulent-like and laminar state, a quasi-periodic heat-flow bursting phenomenon is observed.
		\end{abstract}	
	\maketitle
	\end{CJK*}
	\section{Introduction}
	Natural thermal convection in enclosures filled with fluids has a pivotal role in a long list of engineering and geophysical systems. The fluid motion generated by heating a closed container from below and cooling it above, known as Rayleigh-B$\rm \acute{e}$nard convection (RBC), is a classical model system in fluid mechanics that has been studied over several decades \cite{Castaing1989Scaling,Siggia1994,Ahlers2009,Lohse2010,Chill??2012}. Another frequently encountered natural thermal convection setting, termed vertical convection (VC) system, consists instead of a container heated and cooled from the lateral vertical sides, and it has been equally extensively studied  \cite{Bejan1978Laminar,Yu2007,Patterson2006Unsteady,VC2016_olga,ng_ooi_lohse_chung_2015}. The sole difference between RBC and VC systems is the different directions between the gravity field and the externally imposed temperature gradient, \textit{i.e.}, gravity is antiparallel to the temperature gradient in RBC while it is orthogonal to it in VC. This is responsible for the different stability properties of the two systems, while in the RBC a precise minimal (vertical) thermal gap is necessary for the onset of fluid motion, in the VC a flow always exists independently of the (horizontal) thermal gap.
For a specific geometry, grossly described by the width over height aspect ratio, $\Gamma=L/H$, the internal flow of thermal convection in both such systems is determined by two control parameters, the Rayleigh number $\textrm{Ra}=\alpha g \Delta TH^3/\nu\kappa$ and the Prandtl number $\rm Pr=\nu/\kappa$. Here $\alpha$ refers to the volumetric thermal expansion coefficient, $g$ the acceleration due to gravity, $\Delta T$ the temperature difference between hot and cold boundaries, $\nu$ the kinematic viscosity and $\kappa$ the thermal diffusivity of the fluid. Note that this is true only for small temperature gaps so that the material properties can be reasonably taken as constant, \textit{i.e.}, in Oberbeck-Boussinesq conditions  \cite{Ahlers2009}. One of the key response parameters of thermal convection is the Nusselt number, $\rm Nu$, which provides the ratio of the global mean heat flux through the system over the purely conductive flux. Another key response is the global intensity of turbulence, which can be characterized by the Reynolds number, $\rm Re$, in terms of a suitably defined characteristic velocity for the flow.\\
\indent The dependence of the two response global parameters ($\rm Nu$, $\rm Re$) as functions of the control ones ($\rm Ra$,$\rm Pr$,$\Gamma$) exhibits different behaviors for RBC and VC. Experimental measurements and numerical simulations of the RBC system over wide parameter ranges have revealed many features of the heat transfer mechanisms, and a phenomenological theory capable to predict the functional dependence $\rm Nu(Ra,Pr)$ and $\rm Re(Ra,Pr)$ in the RBC system is nowadays available \cite{GL2001,grossmann_lohse_2000}. In RBC, the exponent of the power-law scaling of $\rm Nu$ on $\rm Ra$ is most of the time in the range between 1/4 and 1/3  \cite{Castaing1989Scaling,GL2001,Ahlers2009,stevens_lohse_verzicco_2011} (it is expected to be larger only in the asymptotic Ra limit \cite{Ahlers2009}). 
In the VC system the same scaling exponent is also found to vary in the same range  \cite{CHURCHILL1975,TSUJI19881723,VERSTEEGH1999,Yu2007,KIS20122625,GEORGE1979813,VC2016_olga} and attempts to frame this into a phenomenological theory have been put forward \cite{ng_ooi_lohse_chung_2015}. However, the VC and the RBC flows can be in different states for the same control parameters. For example, for $\rm Ra\sim O(10^8)$, $\rm Pr\simeq1$ and cylindrical container of aspect ratio 1, the VC flow is in laminar state, while the RBC flow is  turbulent \cite{Shishkina2016a}. This has led to formulate the hypothesis that the presence of a turbulent flow might not be essential in determining the classical near to $1/3$ power-law scaling of $\rm Nu$ as a function of $\rm Ra$ in such systems \cite{Yu2007}.\\
\indent More recently an extension of the RBC and VC system has been studied which takes into account the effect of an arbitrary inclination angle ($\beta$) of the bottom wall of the system with respect to the horizontal, so that the limiting cases $\beta=0^o$ corresponds to RBC and $\beta=90^o$ to VC. Such a system, denoted as confined inclined convection (CIC)  \cite{Shishkina2016a,zwirner_shishkina_2018} has a larger space of control parameters, namely $\rm Ra$,$\rm Pr$,$\Gamma$,$\beta$, and its key response functions Nu and Re appear to be more intricately dependent on the set of control parameters. 
There are presently few series of experimental studies which have explored CIC in the full $[0^o,90^o]$ inclination angle interval. These have been performed either in a quasi-two-dimensional rectangular cell with $\Gamma_x=1$, $\Gamma_z=1/4$ with water, $\rm Pr\simeq6.3$ and $\rm Ra\simeq4.4\times10^9$ \cite{Guo2014} and with silicon oil, $\rm Pr \simeq 10$ and $\rm Ra\simeq6.8 \times 10^{10}$ \cite{Guo2017Evolution}, or in cylindrical cells $\Gamma= 1$ and $\Gamma= 1/20$ with liquid sodium, $\rm Pr=0.0094$, at respectively $\rm Ra=1.47 \times10^7$ \cite{Khalilov2018Thermal} and $\rm Ra = 2.4 \times 10^6$ \cite{Vasiliev2015}. While the first studies (water and silicon-oil) revealed a weak monotonous decrease of the heat flux, with a maximal reduction of $20\%$ with respect to the RBC value, the second group of studies displayed always an overall increase of Nu with the existence of an optimal inclination around $\beta \simeq 65^o -70^o$ for which the increment was $20 \%$ in the $\Gamma= 1$, and remarkably as high as $1100 \%$ in the $\Gamma= 1/20$ system. On the numerical side a thorough series of studies in $\Gamma=1$ cylindrical setting have been conducted for $\rm 0.1\leq Pr\leq100$ and $\rm  10^6\leq Ra\leq10^8$ \cite{Shishkina2016a}. This highlighted  that the relative variation of Nu (with respect to the RBC case) is always moderate $< 25\%$. Furthermore, $\rm Nu(\beta)$  appears to be a non-monotonic function characterized either by one or two local maxima \cite{Shishkina2016a}. A second series of numerical studies have been conducted at $\rm Pr=0.1$ and $\rm Ra$ ranging from $10^6$ to $10^9$ in a narrower  cylindrical system ($\Gamma=1/5$) which showed that Nu increases up to $120\%$ with the inclination angle \cite{zwirner_shishkina_2018}. It is worth mentioning also the only detailed study that reports in parallel experimental and numerical results \cite{Bairi2007}. Such experiments were performed with air, $\rm Pr\simeq 0.7$, at $\rm Ra \leq 10^8$ in a rectangular cell $\Gamma_x=0.75$ and $\Gamma_x=1.5$ and reported too a weak decrease of the relative Nu with the tilting angle.
From the above  mentioned studies one can extract the following general trends:  in unit aspect ratio cells the variation in the global heat-flux with Nu are moderate $< 20\%$. Conversely the variations are much bigger in elongated systems. The form of $\rm Nu(\beta)$ is highly variable (monotonically decreasing, increasing with one or more peaks) and appears to be more sensitive to $\rm Pr$ number rather than to $\rm Ra$.
It appears that the research on the CIC system is still in a taxonomical phase: the current studies indeed try to collect information in order to produce a more complete map of the phenomenology of such a complex system, which depends on at least 4 control parameters (if not more because inclined convection might indeed critically depend also on the cell shape e.g. rectangular or cylindrical). \\
\indent
With this in mind, in this paper we consider the still unexplored case of very large Prandtl number inclined convection cell. More precisely we focus on the dynamics of a quasi-two-dimensional rectangular cell, the same geometry as  \cite{Guo2014,Guo2017Evolution} in a fluid with Prandtl number ($\rm Pr\simeq O(10^2)$) and at Rayleigh numbers ($\rm Ra\simeq O(10^8)$) and as before we control the flow from the developed turbulent state to the laminar state by inclining the system with $\beta\in [0^o,90^o]$. To complement the experiments and explain the observed trends, numerical simulations in very similar conditions are also carried out. We do not seek here to explore the scaling laws of the response parameters as a function of $\rm Ra$ or $\rm Pr$ but rather we focus on the amplitude variation of the global response parameters (together with their temporal fluctuations) as a function of the system inclination angle. Similarly to the question raised in \cite{Yu2007} we aim at understanding whether the presence of a turbulent flow is essential in determining the magnitude of heat transfer in such a system. It appears that, as seen in similar CIC experiments at $\Gamma = 1$ the system self-organize in such a way to robustly keep a stable average global heat flow magnitude, within $20\%$, independently of the inclination angle and so independently of the presence of turbulent fluctuations. 
The functional shape of the modulation of Nu with the inclination is instead peculiar of the studied system, and the behaviour observed here can be in part attributed  to the high flow-temperature coherence displayed by high-Pr fluid dynamics.
		\begin{figure}[h]
	\begin{center}
		\includegraphics[width=70mm]{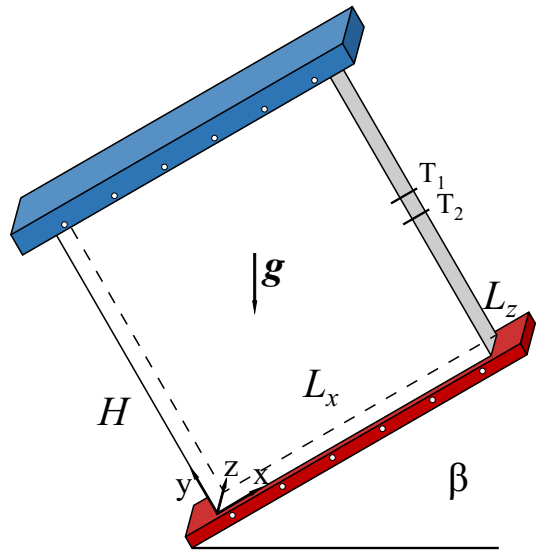}
		\caption{A sketch of the rectangular convection cell and the coordinate frame adopted in this study. The height, length and width of the cell are denoted by $H$, $L_x$ and $L_z$ respectively. $\beta$ is the inclination angle relative to the horizontal plane. The $x$ and $y$ coordinate axes align with the bottom plate and the sidewall respectively. Circles in the top bottom plates indicate the position of the 12 thermistors used to monitor the global thermal gap $\Delta T$, while $\rm T_{1,2}$ show the positions of the two thermal probes employed for the measurement of the intensity of the local flow.}
		\label{Fig1Cell}
	\end{center}
	\end{figure}
	\begin{figure*}[tb]
	\includegraphics[width=180mm]{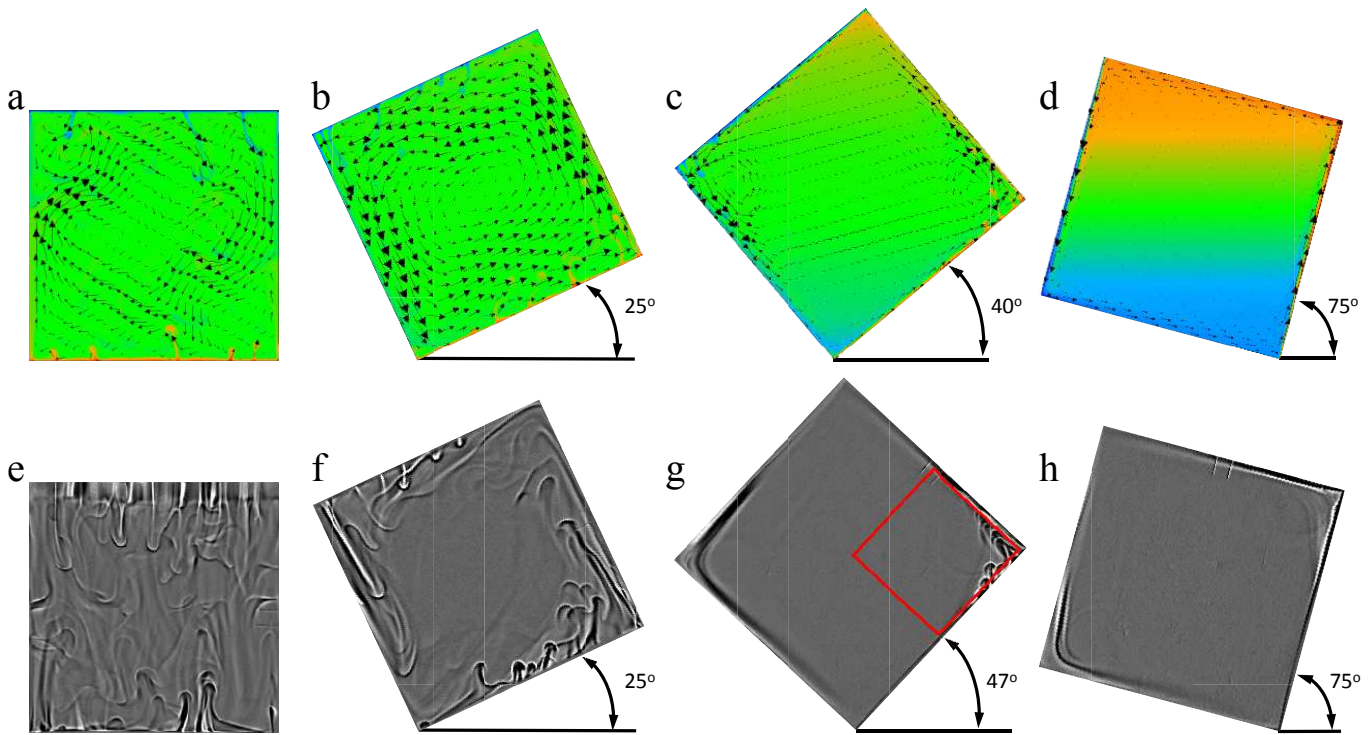}
	\caption{Visualizations of the instantaneous flow pattern in the CIC cell at $\rm Ra=5\times10^8$ from numerics and experiments for various inclination angles.
	Panels of the first raw (a-d) report visualizations both of the velocity and the temperature field in false colours (red-hot, green-intermediate, blue-cold) for the angles 
	 $\beta=0^o$ (a),  $\beta=25^o$ (b),  $\beta=40^o$ (c),  $\beta=75^o$ (d). The second raw (e-h) shows the shadowgraph images of a vertical central plane in the CIC system from the experiments at  $\beta=0^o$ (e),  $\beta=25^o$ (f),  $\beta=47^o$ (g),  $\beta=75^o$ (h).}
	\label{snap}
	\end{figure*}	
\section{Setups and methods}
\subsection{Experiments}
The experimental setup employed in this study is sketched in Fig.~\ref{Fig1Cell}. It is a rectangular cell, with height $H = 240\ mm$, length $\ L_x = 240\ mm$ and width $\ L_z = 60\ mm$. The relative aspect ratios of the length and width with respect to height are therefore $\Gamma_x=L_x/H$$=1$ and $\Gamma_z=L_z/H$$=1/4$ respectively. The experiments presented here have been conducted by means of silicone oil as working fluid at a mean temperature $T_m= 25$$\rm ^\circ C$ which corresponds to a Prandtl number $\rm Pr\simeq480$. The lateral walls of the cell are made by acrylic foils of $5\ mm$ thickness and they are kept within two thick copper top and bottom plates of 35mm and 25mm respectively supported by four steel pillars. 
The temperature of the top plate is kept constant by a refrigerated liquid circulating bath (Polyscience AD15R-40-A12Y) with a temperature stability of $\pm0.01$ K. While a film heater is embedded into the bottom plate and it is powered by a power supply 
in order to provide a constant heat flux. In order to decrease the systematic errors associated to possible heat losses a thermostat box, with setting at ambient temperature at $\rm 25^\circ C\pm0.4^\circ C$, is used to shield the full convection system. 
The temperatures of the plates are measured by means of six embedded thermistors per plate, whose electric resistances are measured by a digital multimeter (NI PXIe-4065) at a constant sampling rate of 3 Hz. The corresponding temperatures are calculated via the Steihart-Hart function \cite{STEINHART1968}.
The experiments are performed at two time-averaged temperature differences, $\Delta T$=$4.0\pm0.1$ K (corresponding $\rm Ra\simeq10^8$) and $\Delta T$=$20.0\pm0.1$ K (corresponding $\rm Ra\simeq5\times10^8$). With such temperature differences non-Oberbeck-Boussinesq effects in silicon oil are negligible. 

The measurements of the Nusselt number in the experiments is based on the relation $\rm Nu = $$Q H/ (\chi \Delta T)$, where $Q$ is the time-averaged total heat-flux through the system and $\chi$ the thermal conductivity of the fluid. Here $Q$ is calculated from the electric power used by the heater ($P$) via the relation $Q= P/(L_x\ L_z) = V^2/ (R\ L_x\ L_z)$, where $V$ is the voltage, $R$ the electric resistance of the heater. The measurement of the Reynolds number, $\rm Re =$$ U H /\nu$, requires instead an estimate of a characteristic velocity amplitude of the flow, $U$. In the experiment we take for $U$ the local fluid velocity in a position at half height in the cell and $1\ cm$ away from the sidewall, as shown in Fig.~\ref{Fig1Cell}. This is done by means of two thermistors ($T_1$ and $T_2$ in Fig.~\ref{Fig1Cell}) of 0.2 mm in diameter and $1.5\ cm$ apart. We use two Wheatstone bridges and lock-in amplifiers to measure the temperatures from such probes at a sampling rate of 128 Hz. At this sampling rate the cross-correlation from the thermistor time-series results to be smooth and the elliptic model  \cite{He2006Elliptic} can be used to reconstruct the intensity of the velocity.

\subsection{Numerical simulations}
Next to the experiments we conduct direct numerical simulations (DNS) of thermal convection in an inclined cell at the same $\rm Pr$ and $\rm Ra$ numbers as in the experiments. We adopt however the simplifying assumption of i) having a two-dimensional system ($\Gamma_x=1$ and $\Gamma_z = 0$), ii) using the Oberbeck-Boussinesq approximation and finally iii) specifying constant temperatures boundary conditions (BC) on top-bottom plates, while for the lateral ones we assume perfect insulation conditions. Note that no-slip BC for velocity is taken on every domain boundary. This means that the effects associated to the flow movement in the transverse direction (z-axis), or to possible non-linear buoyancy effects \cite{Ahlers2008Non}, or to the top-bottom asymmetries in the thermal BC which are inevitably present in the experiments  \cite{Huang2015A,PhysRevLett.102.064501}, are all ignored in the calculations. The simulations are performed by means of a multi-population single-relaxation Lattice Boltzmann solver, the same as in Ref.~\cite{Shrestha2016Finite}; the computational grids of simulations - which are regular and cartesian - have 512 $\times$ 512 nodes or 1024$\times$1024 nodes for the more demanding cases in spatial resolution which occur for $\rm Ra=5\times10^8$ at large inclination angles. In order to obtain statistically converged global quantities each combination of $\rm Ra$ and inclination angle $\beta$ is simulated for more than 100 large scale circulation times, $\tau = H/u_{rms}$ in statistically stationary conditions. The Nusselt number in simulations is computed via $\rm Nu$=$( \left\langle u_y T\right\rangle -\kappa \partial_y\left\langle T\right\rangle) /\left( \kappa\Delta T/H\right) $, where $\langle \ldots \rangle$ represents an average over time and volume. We will indicate with $Nu_t$ the instantaneous value of the Nusselt number, which is computed by replacing $\langle \ldots \rangle$ by the volume average $\langle \ldots \rangle_V$. 
\section{Flow patterns}
We begin by characterizing from a qualitative perspective the different flow regimes observed in the CIC system at varying its inclination angle. 
Two recurrent flow and temperature spatially coherent structures are found in convective cells and particularly in the RBC system: the thermal plumes (TP) and the large scale circulation (LSC), both of them have significant effects on the global heat transfer in the system \cite{Shang2003,Shang2004,Funfschilling2004}. The plumes are hot or cold region of the flow that detach in an apparent random fashion from the horizontal thermal boundary layers (BL) and that impinge rapidly into the bulk of the cell under the effect of buoyancy. Their thickness is comparable to the one of the thermal BL but they can lose quickly their coherence under the effect of diffusion or by the turbulent mixing which occurs in the bulk of the cell. The LSC is on the contrary a roll-like flow movement that fills in the full cell producing a wind parallel to all the walls in the system. In the RBC the self-organized LSC has been observed to be either stable or bistable or absent depending on the Pr and Ra values of the system  \cite{Brown2006Rotations,Sugiyama2011Flow,Araujo2005Wind}, while it is always present and oriented in the same direction in the VC \cite{Guo2014,Shishkina2016a}. We also note that a stronger LSC is known to promote the plume detachment, however the contribution of LSC to the total heat flux with respect to the one of TP is found to be negligible in the RBC for $\rm Ra$ ranging from $10^5$ to $10^{11}$ \cite{Belmonte1993Boundary}. 
	\begin{figure}[h]
		\includegraphics[width=80mm]{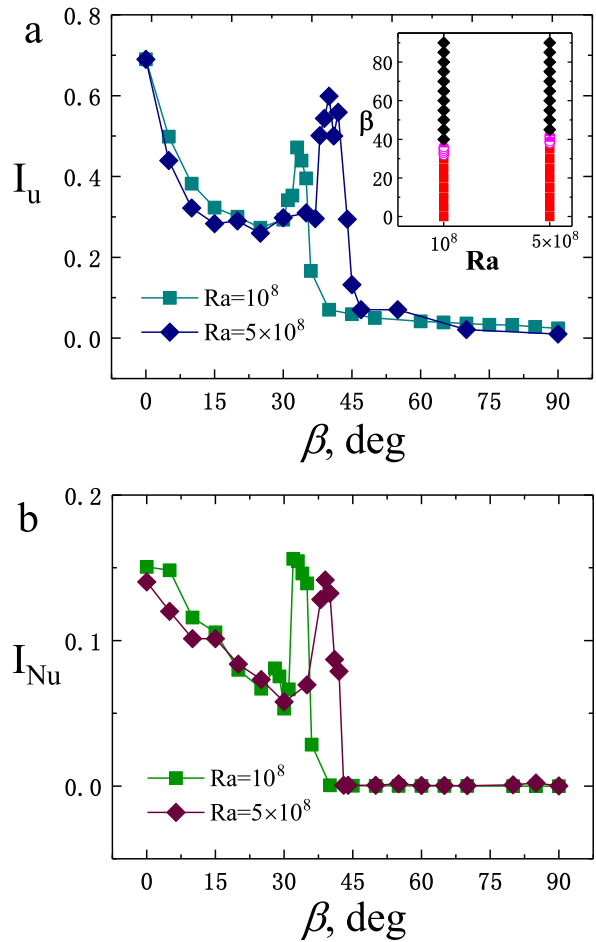}
		\caption{(a) The intensity of the turbulent flow $I_u(\beta)$ in inclined thermal convection for $\rm Ra=10^8$ (squares) and $\rm Ra=5\times10^8$ (diamonds). Inset: the phase diagram of the flow pattern in the $\rm Ra$ and $\beta$ plane, the black diamonds refer to the laminar regime, the red squares refer to the turbulent regime and the pink circles refer to the bursting phenomenon regime. (b) The fluctuation intensity of the heat transfer $I_{Nu}(\beta)$ in the inclined thermal convection for $\rm Ra=10^8$ (squares) and $\rm Ra=5\times10^8$ (diamonds).}
		\label{I_phase}
	\end{figure}
	\begin{figure}[h]
		\includegraphics[width=80mm]{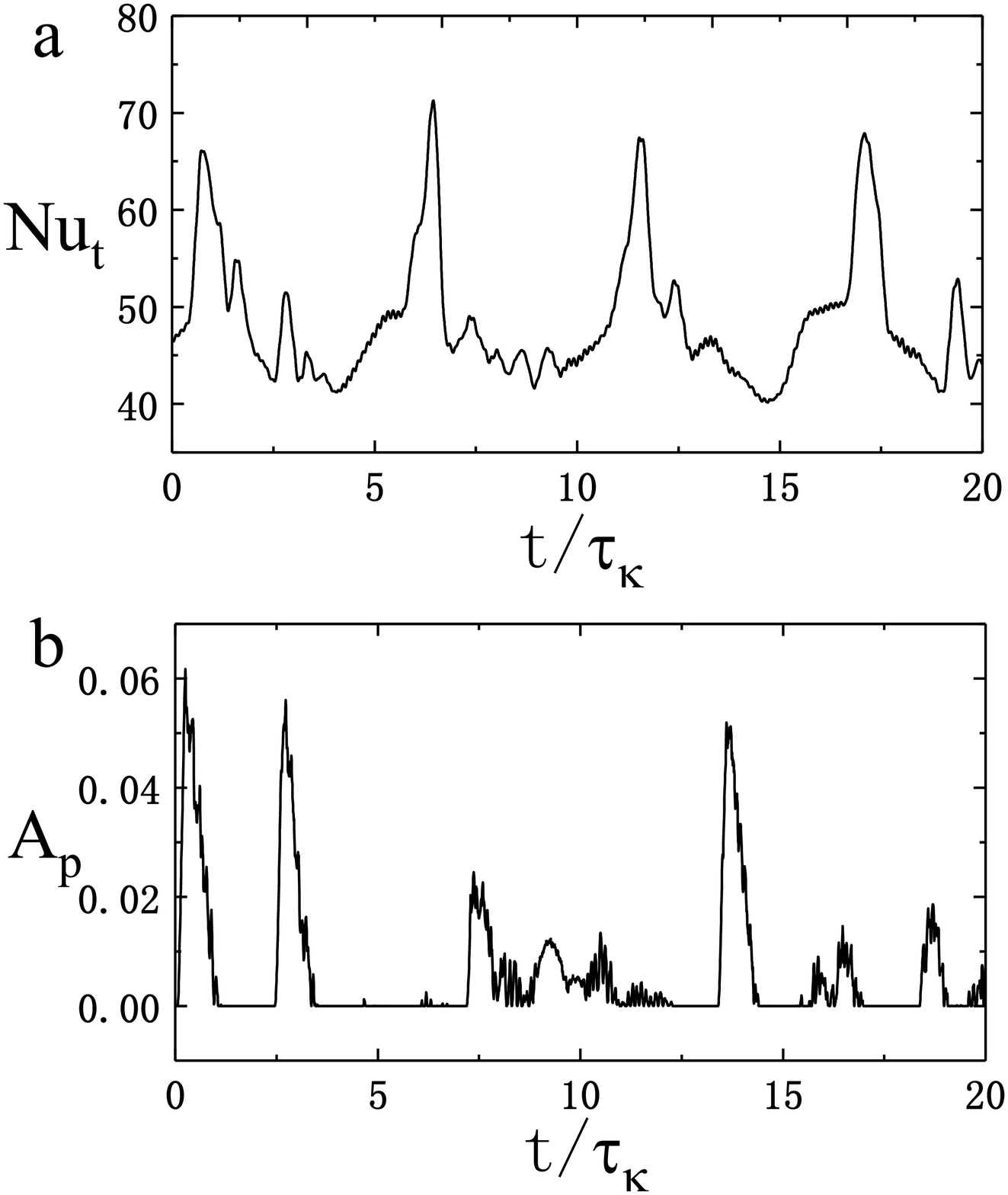}
		\caption{(a) Time series of instantaneous (volume-averaged) Nusselt number at $\beta=42^o$ for $\rm Ra=5\times10^8$ in simulation. The averaged period of the bursting phenomenon in simulation is about 5$\tau$. (b) Time series of the area fraction of plumes $I_p$ at $\beta=47^o$ for $\rm Ra\simeq5\times10^8$ in experiment. The averaged period of the bursting phenomenon is close to $6\tau$.}
		\label{Nu_tau}
	\end{figure}
In Fig.~\ref{snap} we report a series of instantaneous visualizations of the flow in the CIC system at different inclination angles and in statistically stationary conditions. In particular we display next to each other snapshots of the velocity and temperature flow field from numerics (panels a-d) and shadowgraph images - whose shading is roughly proportional to the temperature gradient along z - from experiments (panels e-h) at similar inclination angles. The patterns observed appear to be qualitatively similar, supporting therefore the assumption of a quasi-bidimensional dynamics in our CIC experimental system. In the horizontal case, $\beta=0$ (Fig.~\ref{snap}a,e), the flow is composed by disordered TP emerging from the boundary layers. The LSC structure is weak, and a single roll can not be clearly identified, this is corroborated by the fact that the plumes, both in the experiment and in the DNS, emerge almost vertically from the BL. A similar flow state, denoted as soft turbulence, has been observed in other high-Pr RBC studies \cite{Shang2003,Shang2004,Xi2009Origin}.  Note that the term soft turbulence is used to stress the small scale separation in the flow. Indeed in the RBC, 
the ratio between the global Kolmogorov scale and the systems size, $\eta  /  H$,  can be related exactly to the control parameters $\eta  /  H = Pr^{1/2} (Ra (Nu-1))^{-1/4} $ \cite{Sugiyama2007Glycerol}. It turns out that in our system $\eta  /  H = 0.1$ at $\rm Ra=10^8$ and $\eta  /  H = 0.06$ at $\rm Ra=5 \times 10^8$.
For a moderate inclination angle, $\beta=25^o$ (Fig.~\ref{snap}b,f), the LSC appears instead to be well established, the circulation has an elliptical shape at this angle, and becomes markedly more space filling and square in shape when $\beta$ is increased, which is consistent with Ref.~\cite{Guo2014}. A similar change of LSC shape, from elliptical to square, is also found in the horizontal RBC when Ra (and so Re) is increased \cite{Niemela2003Rayleigh,Xia2003Particle}. At a large inclination angles, $\beta=75^o$ (Fig.~\ref{snap}d,h), the intensity of the velocity results dramatically suppressed.  A LSC is still present but any turbulent feature has disappeared, the flow is regular at the boundaries and nearly quiescent in the bulk. As a result the temperature does not display any plume structure and it appears horizontally stratified except for the regimes near to the boundaries, a finding consistent with the numerical results in \cite{Shishkina2016a}. The latter flow state remains nearly unchanged for the $\beta=90^o$ VC case.\\
\indent A peculiar phenomenon is observed for $\beta \sim 45^o$ (Fig.~\ref{snap}c,g), in correspondence to the transition regime from the irregular turbulent-like flow state to the regular one. In such a state, we observe that the TP get generated only on the corners of the cell where the fluid which moves parallel to a hot/cold wall is constrained by the boundary to change direction and, more importantly, the plumes are generated simultaneously in a nearly-periodic fashion. This bursting behaviour, which to our knowledge as not yet been reported elsewhere, is observed both in experiments and simulations although with a difference: in experiments, due to the asymmetry in the thermal boundary conditions,  the plumes do not always detach from the top and bottom boundaries simultaneously (see Supplemental Material \cite{Supplemental} for movies).  
To better identify the angles at which the bursting phenomenon occurs we investigate, in the DNS,  the behaviour of the normalized temporal fluctuations of the global velocity intensity, which we define as $I_u=\sigma_t(u_{rms})/\langle u_{rms}\rangle_t$ where $u_{rms}(t) = \sqrt{\langle\textbf{u}\cdot\textbf{u}\rangle_V}$ and $\sigma_t(\cdot)$ denotes the temporal standard deviation, and at the same time the fluctuation intensity of heat transfer, $I_{Nu}=\sigma_t {(Nu_t)} / Nu$, with $\rm Nu_t$ the global instantaneous value of the Nusselt number.
Both the measurements of $I_u(\beta)$ and $I_{Nu}(\beta)$, reported in Fig.~\ref{I_phase}, decrease for small inclinations in the turbulent regime and bump-up to the turbulent level in the bursting state, while for larger angles they decrease rapidly to a vanishing value, in support of the observation that the flow becomes laminar and time independent. 
The inset of Fig.~\ref{I_phase} shows the phase diagram of the state of the CIC system in $\rm Ra\times \beta$ space, where we distinguish between the turbulent or chaotic, the bursting and the laminar state, by thresholding over $I_u(\beta)$  (the bursting regime is identified by the criterion $|\partial I_u/\partial \beta|>$1.15). We find that the bursts occur in a narrow angle range: $\beta \in [ 32^\circ, 36^\circ]$ at $\rm Ra=10^8$ and $\beta \in [ 38^\circ, 42^\circ]$ at $\rm Ra= 5\times10^8$.
Finally, in Fig.~\ref{Nu_tau} we report the time series for the instantaneous Nusselt number $\rm Nu_t$ in DNS for an angle where the system is in the bursting regime. It is evident its oscillatory character with peaks up to 40$\%$ higher than the mean. The peaks in Nusselt corresponds to a release of cluster of plumes from the BL into the bulk, proving that in this regime the heat transport by convection is still strongly affected by thermal plumes rather than by the mean flow, in agreement with Ref.~\cite{Shang2003}. 
Furthermore it appears that the time-scale associated to the bursts is of the order of $ 5 \tau_{\kappa}$ where $\tau_{\kappa} = \lambda_T^2/\kappa$ indicates the thermal diffusion time based on the thermal boundary thickness $\lambda_T=H/(2 Nu)$, see again Fig.~\ref{Nu_tau}. Note however that the observed periodicity is also close to the typical turn over time $\tau$ at that inclination, which makes difficult to distinguish the exact physical mechanism for this phenomenon.
In the experiments the bursting regime can be highlighted by looking at the time series of the amount of plumes detected in the shadowgraph movies. In order to identify them we adopt the plumes identification procedure used in Refs.~\cite{Guo2017Evolution,Zhou2006Onset}, which is based on a threshold analysis on the intensity variance of the shadowgraph images, and we then calculate the area fraction of the plumes, $A_p$ appearing in one of the corner of the experiment (inside the red square frame shown in Fig.~\ref{snap}g). Large values of $A_p$ indicates that more plumes are measured in the shadowgraph images. As shown in Fig.~\ref{Nu_tau}b, despite the fact that in the experiments the bursting phenomenon is less regular, it appears to occur with a similar frequency.
	\section{Heat flux variation}
	\begin{figure}[h]
		\includegraphics[width=80mm]{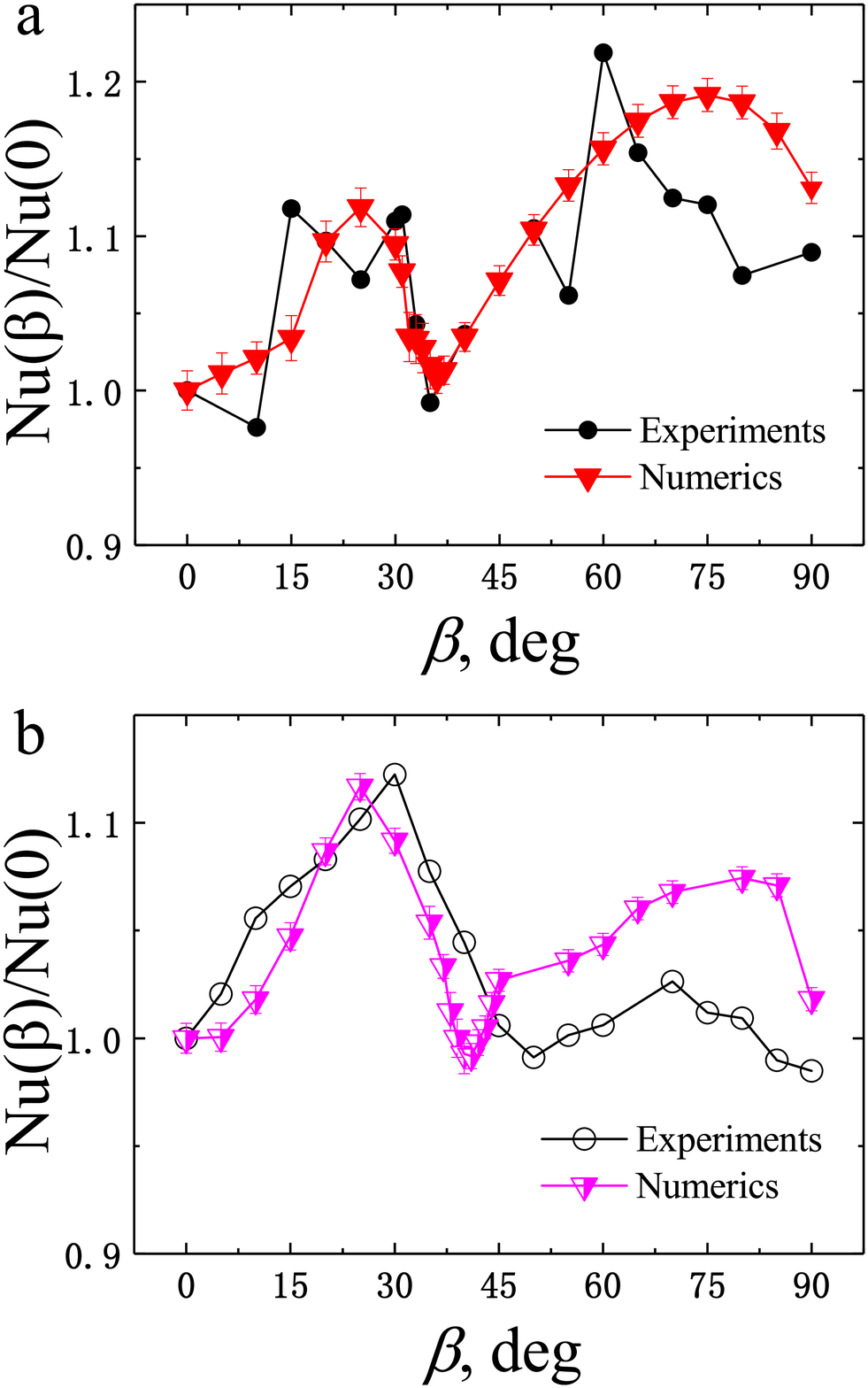}
		\caption{Nusselt number $\rm Nu(\beta)$ as a function of the inclination angle $\beta$, normalized by its value for $\beta=0$, from experiments (circles) and numerics (triangles) at (a) $\rm Ra=10^8$ and (b) $\rm Ra=5\times10^8$. In the numerics the error bars are estimated from $\delta_{\frac{Nu(\beta)}{Nu(0)}}=\frac{\rm Nu(\beta)}{\rm Nu(0)}\sqrt{\left(\frac{\delta Nu(\beta)}{ Nu(\beta)}\right)^2+\left(\frac{\delta Nu(0)}{ Nu(0)}\right)^2}$.}
		\label{exp_sim}
	\end{figure}
We now turn the attention on the global and time averaged heat flux in the system, in other words to the Nusselt number response function. 
When studying the dependence of $\rm Nu(\beta)$ it is convenient to take as a reference the Nusselt value measured in the horizontal RBC system $\rm Nu(0)$.  In the experiments we measured $\rm Nu(0)\simeq31.5$ for $\rm Ra\simeq10^8$ and $\rm Nu(0)\simeq53.8$ for $\rm Ra\simeq5\times10^8$, which is in agreement with the results obtained in a similar setup at $\rm Pr\simeq554$, $\Gamma=1$ \cite{Xia2002Heat}. The corresponding values computed from the simulations are $\rm Nu(0)=29.3 \pm 4.4$ for $\rm Ra=10^8$ and $\rm Nu(0)=48.5 \pm 6.9$ for $\rm Ra=5\times10^8$ and are compatible within error bars. Note that in the simulations the statistical error for $\rm Nu$ are estimated from the temporal standard deviations of $\rm Nu_t$ by means of $\delta Nu = \sigma_t (Nu_t) \sqrt{2\tau_I/\mathcal{T}}$, with $\tau_I$ the integral time for $\rm Nu_t$, and $\mathcal{T}$ the total simulation time \cite{sugiyama_calzavarini_grossmann_lohse_2009}. On the contrary, in the experiments the positioning of the thermistor inside plates of high thermal inertia prevents from correctly measure the temporal fluctuations of the Nusselt number which turns out to be highly suppressed, as a result a reliable statistical error bar of $\rm Nu$ from experiments could not be evaluated.\\
\indent Figure \ref{exp_sim} shows the normalized Nusselt number $\rm Nu(\beta)/Nu(0)$ both from experiments and from numerics for the two  studied Ra numbers. 
First, we note the quantitatively similar behaviour between experiments and simulations, which is remarkable for the higher-Ra case. Second, we notice that the relative variations of Nu are small: for instance in the DNS the increases of Nu are close to 11.8$\%$ for $\rm Ra=10^8$ and 11.6$\%$ for $\rm Ra=5\times10^8$ in the turbulent dominated regime, while in the laminar regime, the enhancements of Nu reaches 19$\%$ for $\rm Ra=10^8$ and 7.4$\%$ for $\rm Ra=5\times10^8$. In the experiments the trend are very similar with overall slightly weaker enhancements in the laminar regime as compared to DNS.  The little variations of Nu as a function of the inclination angle is a remarkable effect if one considers that at varying the angle the nature of the flow changes from a turbulent and plume dominated one to a time independent laminar flow and it attests for what we denote as robustness of the heat transfer convective process in this type of system. 
Looking more in detail on the $\rm Nu(\beta)$ trend one can observe a ``bimodal" dependence, characterized by the presence of two local-maxima.
For $\rm Ra=10^8$ in simulations, the Nu increases in the turbulent dominated regime to a local maximum at $\beta \sim 25^o$ and descends to a minimum point at $\beta=36^o$ in the bursting flow regime. For higher angles Nu increases again to a second maximum at $\beta \sim 75^o$ and drops a little for $90^o$. For $\rm Ra=10^8$ in experiments, the bimodal trend is less distinct, but the agreement on Nu amplitude is clear in the bursting regime. The curves of $\rm Nu(\beta)/Nu(0)$ for $\rm Ra=5\times10^8$ have a similar trend although with a tiny shift towards larger angles for the local minima corresponding to angle shift in the bursting regime.

We remark that these results are significantly different from the ones of previous experiments in the same cell geometry at smaller Prandtl number $\rm Pr\simeq6.3, \rm Ra \sim10^9$ \cite{Guo2014}  and $\rm Pr\simeq10, \rm Ra \sim 10^{10}$ \cite{Guo2017Evolution} in which $\rm Nu$ monotonously decreases by 20$\%$ by increasing $\beta$. Instead a qualitatively similar double-peaked variation of $\rm Nu$ is found in DNS in a cylinder of aspect ratio 1 at $\rm Pr=10$ and $\rm Pr =100$, $\rm Ra=10^7$ \cite{Shishkina2016a}. The same numerical study shows that evolution of $\rm Nu(\beta)$ at changing Pr for fixed Ra evolves from the one with single-peak to one with two peaks. This points to the fact that the magnitude of $\rm Pr$ may be of key importance in this respect.
\section{Reynolds number variation}
To complement our findings on the heat flux we investigate the variation of the response function given by the Reynolds number at varying the inclination angle.
As mentioned before, one way to define the Reynolds number is by means of the global root-mean-squared velocity, $u_{rms} = \sqrt{\langle\textbf{u}\cdot\textbf{u}\rangle}$ so $\textrm{Re} = u_{rms} H/\nu$, however in the simulations, where the full velocity field is available, it is convenient to consider also Reynolds number based on the single $rms$ components of the velocities, \textit{i.e.}, $\rm Re_x$ and $\rm Re_y$. On the contrary, in the experiment the measurement of the velocity field is only performed locally by analyzing the cross correlation signal from a couple of thermistors positioned next to an insulating wall in the cell. Such a method works only when the mean flow is robust \cite{He2006Elliptic} (see also Refs.~\cite{He2015Reynolds,zhou_li_lu_liu_2011} for its application in the RBC system), and in conclusion it may be applied only when a vigorous LSC is present in the system. This limitation prevented us from measuring the velocity field for every inclination, and made it possible to have a reliable Reynolds estimate, denoted $\rm Re_l=$$U_yH/\nu$ (where $U_y$ denotes the local y direction rms velocity estimated by elliptic model \cite{He2006Elliptic}), only for the higher Ra number experiment. 
All the described Reynolds number measurements are reported as a function of the tilting angle $\beta$ in the main panels of Fig.~\ref{Re_Re}(a,b). Their intensity if we compare the two Ra varies of about a factor $\sqrt{5}$ in a agreement with the approximate scaling law $\rm Re \sim Ra^{1/2}$ \cite{Grossmann2000Scaling}, which is larger than the variation detected in the amplitude of $\rm Nu$ (which scales roughly as $\rm \sim Ra^{1/3}$ or $\rm \sim Ra^{1/4}$). 
Also note that the magnitude of Re for $\beta = 0^{\circ}$ is much smaller than one observed in  RBC at small Pr \cite{zwirner_shishkina_2018}.
For what concerns the $\beta$ dependency, we observe that in the turbulence dominated regime all the curves increase to a maximum corresponding to a weak tilting at $\beta\simeq10^\circ$ and then they drop rapidly till a value $\rm Re \sim 2$ for angles which correspond roughly to the occurrence of a local minimum in the $\rm Nu(\beta)$ (see again of Fig.~\ref{exp_sim}). Finally, for larger angles the decrease in $\rm Re$ is much more gentle and reaches values which correspond to a creeping flow. 
In Fig.~\ref{Re_Re}(b) we see that the single peak behaviour for $\rm Re(\beta)$ is also supported by the measurements of $\rm Re_l$ from the experiment, which are in good agreement with the numerical value of $\rm Re_{x,y}$. Notice however that the $\rm Re_l$ at large $\beta$ could not be estimated in the experiments for the laminar regime due to the extreme weakness of the flow.
Let us remark that the curves of Re is again different from the experimental results at smaller Prandtl number, where it is generally observed a peak occurring also at $\beta \simeq 10 \pm 2^o$ but followed by a linear decrease till the largest inclinations \cite{Guo2014,Guo2017Evolution,Shishkina2016a}
	\textsc{\begin{figure}[h]
			\includegraphics[width =85mm]{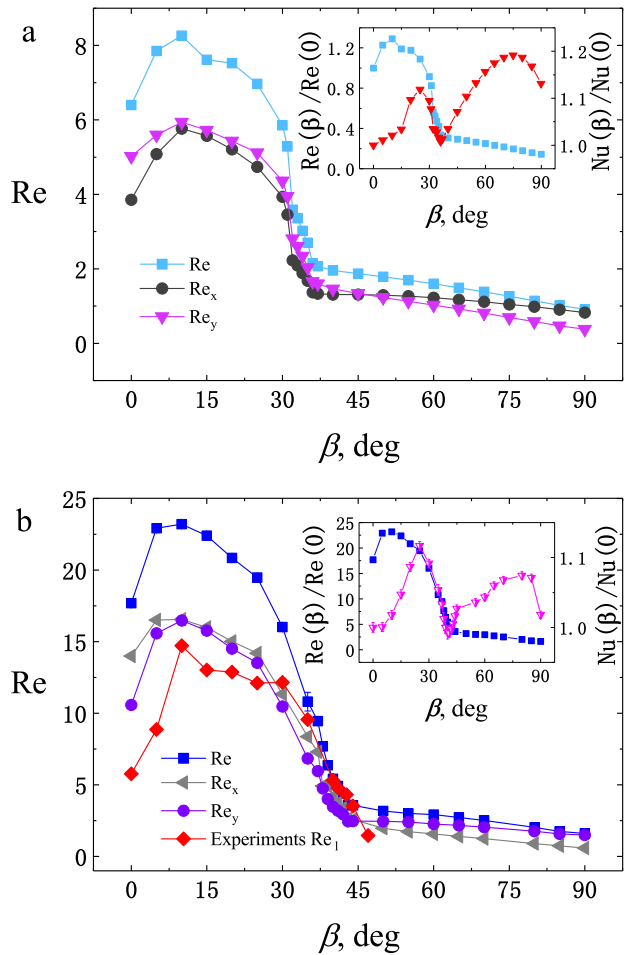}
			\caption{Reynolds numbers as functions of the inclination angle $\beta$: (a) for $\rm Ra=10^8$. (b) for $\rm Ra=5\times10^8$. In (b) we report also the local Reynolds number $\rm Re_l$ from experiments is (red diamonds). In the insets: the superposed graphs for $\rm Re(\beta)/Re(0)$ (squares) and $\rm Nu(\beta)/Nu(0)$ (triangles) from DNS.} 
			\label{Re_Re}
	\end{figure}}
	\begin{figure}[h]
		\includegraphics[width=85mm]{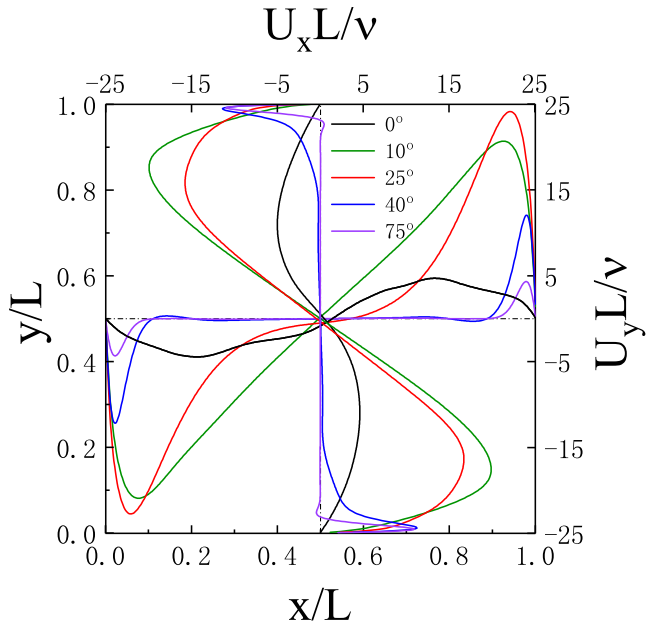}
		\caption{The x-axis and y-axis velocity profiles $U_x(y)$ and $U_y(x)$ as derived from the conditionally time averaged velocity components $u_x$ and $u_y$ at different inclination angles for $\rm Ra=5\times10^8$.
		}
		\label{profiles}
	\end{figure}

To better appreciate the link between the behaviours of $\rm Re$ and $\rm Nu$ at varying the inclination we plot, in the insets of Fig.~\ref{Re_Re}, their normalized trends with respect to the flat cell case. First let us observe that the overall variations of $\rm Re(\beta)/Re(0)$ are of more than $85 \%$, much more pronounced than for $\rm Nu(\beta)/Nu(0)$. 
Furthermore, in the turbulent regime even if the peaks in $\rm Re$ and $\rm Nu$ are not coincident it appears that after such maxima there is an evident correlation between the sharp decrease of both these observables. In the laminar regime however, such correlation is lost and the robustness of the heat-flux seems therefore not connected to fluctuations of the velocity.
\begin{figure}[h]
	\includegraphics[width =85mm]{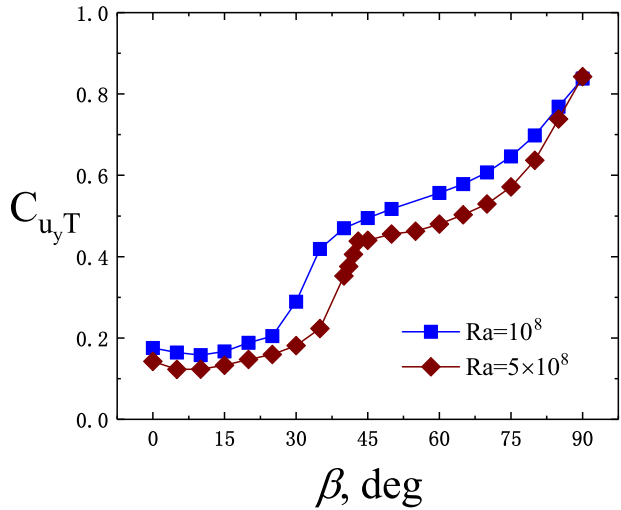}
	\caption{Correlation coefficient of $y$-axis velocity $u_y$ and temperature $T$, $C_{u_y T}$, in inclined convection as a function of inclination angle $\beta$ for $\rm Ra=10^8$ (squares) and $\rm Ra=5\times10^8$ (dimands).}
	\label{Nu_Co}
\end{figure}

The non-coincidence of the peaks in $\rm Re(\beta)$ and $\rm Nu(\beta)$ may be due to the fact that the definition of the $\rm Re$ based on the $rms$ velocity does not reliably account for the the intensity of the LSC. In order to properly quantify the intensity of the LSC, rather than measuring the velocity fluctuations one should extract the amplitude of the mean flow in the system. To this aim we apply to the numerical data a technique proposed in \cite{sugiyama_calzavarini_grossmann_lohse_2009}, based on conditional average of the flow field with respect to the sign of the vorticity, $\omega$, in a point at the centre of the system which allows to compute both time averaged velocity flow patterns $\langle u_x\  \rm sign(\omega) \rangle_t $  and $\langle u_y \ \rm sign(\omega) \rangle_t$.
In Fig.~\ref{profiles} we show the average spatial profiles of these conditional fields, i.e., $U_x (y) \equiv  \langle \langle u_x\  \rm sign(\omega) \rangle_t \rangle_{x(y)}$ and $U_y(x) \equiv \langle \langle u_y \ \rm sign(\omega) \rangle_t \rangle_{y(x)}$, computed over a time history of about 20 $\tau$ and where $\langle \ldots \rangle_{x(y)}$ and $\langle \ldots \rangle_{y(x)}$ represent the line averaging along the x direction for fixed y or along the y direction for fixed x.
This proves that the first peak of $\rm Nu(\beta)/Nu(0)$ is due to the enhancement of the LSC (and specifically the enhancement of its $y$-axis component).
From Fig.~\ref{profiles} we also see that a LSC is still present in the laminar regime, however it decreases by increasing $\beta$ and becomes progressively more attached to the walls (one expects indeed that the thickness of a laminar boundary layer will be proportional to $\rm Re_{x,y}^{1/2}$). It is not possible in this case 
to correlate the $\rm Re(\beta)$ trend with the second peak observed in $\rm Nu(\beta)$.
	\section{Heat-flux and coherent flow structures}
The fact that the intensity of the heat flux remains high even in the laminar regime where the mean flow is progressively weakening and the temporal fluctuations are absent is a puzzling effect. In order to make sense on this phenomenon it is useful to look at the time-averaged global correlation coefficient of the velocity component in the direction of the imposed thermal gradient, $u_y$, with the temperature field itself, which is defined as:	
		

		\begin{equation}
		C_{u_y T}=\left\langle\frac{\left\langle u_y' T'\right\rangle_V }{\sqrt{\langle u_y'^2 \rangle_V \langle T'^2 \rangle_V}}\right\rangle_t
		\end{equation}
		where $u_y' = u_y - \langle u_y \rangle_V$ and $T' = T - \langle T \rangle_V$ denotes the deviation from the mean, note however that $\langle u_y \rangle_V=0$ because the system is bounded and $ \langle T \rangle_V = (T_{hot} + T_{cold})/2$ the arithmetic mean of the plates temperatures.
The function $C_{u_y T} (\beta)$ represents the efficiency of the flow in transporting the heat across the cell, and it is shown in Fig.~\ref{Nu_Co} for both Ra of the numerical study.  
In the turbulent dominated regime, $C_{u_y T}$ is small because the only regime where the velocity and temperature fields are correlated is in the TP, which are destroyed by the turbulent fluctuations of the bulk and so they occupy a small fraction of the whole system, furthermore the overall velocity intensity which is in the denominator is large as so the resulting overall correlation is small. In the bursting regime, $C_{u_y T}$ begins to increase rapidly, this due to the decrease of the velocity intensity, and finally becomes very large in the laminar regime, indicating that the residual slow motion of fluid becomes more effective in transporting heat. Remarkably, for the largest inclination which corresponds to the VC system the correlation is as high as $85\%$. 

\begin{figure}[h]
		\includegraphics[width =85mm]{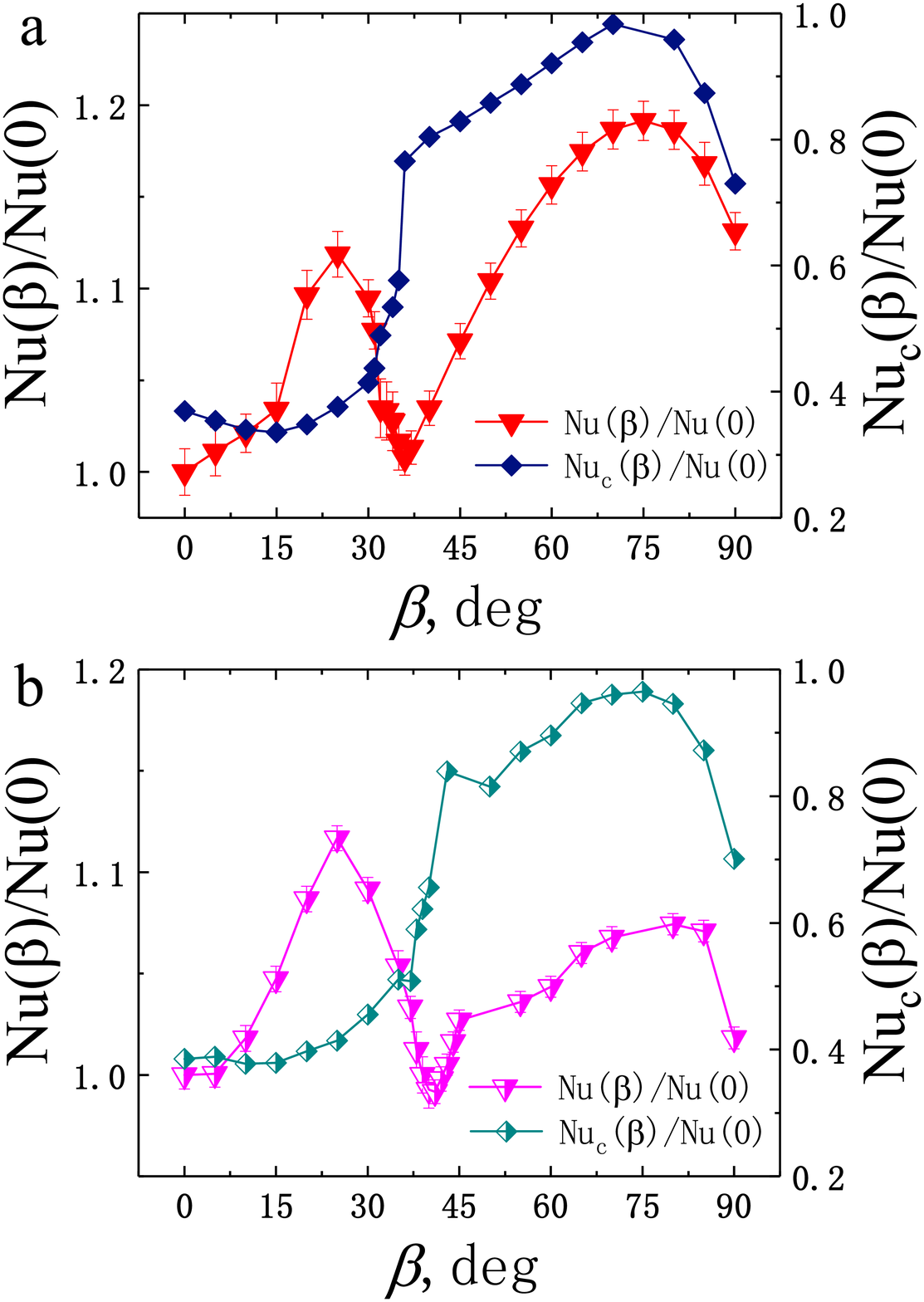}
		\caption{Normalized Nusselt number $\rm Nu(\beta)/Nu(0)$ (triangles) and fraction of heat transfer, $\rm Nu_c(\beta)/Nu(0)$, carried by the coherent structure (diamonds) for (a) $\rm Ra=10^8$ and (b) $\rm Ra=5\times10^8$ as functions of the inclination angle $\beta$.}
		\label{Nuc}
	\end{figure}

Such an analysis can be further refined by quantifying the amount of heat $\rm Nu_c$ which is transported through the coherent structures.  The method we adopt here was first proposed for the RBC system in order to identify the plumes in a convective flow \cite{Ching2004Extraction,Van2015Plume,Huang2013Confinement}. This method is based on a thresholding procedure both on the instantaneous correlation field of $u_y$ with $T$ and on the instantaneous intensity of the temperature itself (in accordance with Ref.~\cite{Van2015Plume}, the empirical thresholding constant on the correlation $c$ is chosen to be 1.2). 
The resulting visualizations (not shown here) identify as coherent flow regions mostly the TP in the turbulent regime and the thin LSC roll in the laminar one.
Though the magnitude of $\rm Nu_c(\beta)/Nu(0)$ slightly depends on the criterion of detection of coherent structure, it is clear from Fig.~\ref{Nuc} that there exist a strong correlation between $\rm Nu(\beta)/Nu(0)$ and $\rm Nu_c(\beta)/Nu(0)$  only in the laminar regime. Therefore, it appears even more clearly that the large heat flow observed for large inclinations (laminar regime) is due to the strong correlation of temperature and velocity in the LSC region of the flow. While, on the contrary in the turbulent regime such an identification between flow structures (LSC or TP) and heat transport is much more complex, or at least not detectable with this analytical tool.
Note that the value of the $\rm Pr$ number plays a crucial role here. Indeed the time it takes for a patch to be spread by diffusion with respect to the time it takes to be transported by a flow of given intensity is given by the P\'eclet (Pe) number that is $\rm Pe  = Re \ Pr$. This means that even when the flow is in creeping conditions, take e.g. $\rm Re \sim 0.2 $ the convective time is $O(10^2)$ shorter than the diffusion one ($\rm Pe \sim 100$). This might offer a key on why such a trend for $\rm Nu(\beta)$ in the laminar regime  was not observed in any of the studies at lower-Pr.
\section{Conclusions}
In this paper, we have investigated the behaviour of confined inclined convection in a quasi-bidimensional rectangular system of unit aspect ratio at a constant Prandtl number, $\rm Pr\simeq480$, and Rayleigh numbers ($\rm Ra=10^8$ and $5\times10^8$) by means of experiments and simulations. In particular we have focused on the characterization of the flow patterns and on the key response functions $\rm Nu$ and $\rm Re$ with respect to the inclination angle.  By progressively tilting the system from $\beta=0^o$ to $\beta=90^o$  the flow changed from the a soft-turbulent state, dominated by plumes which penetrate well into the bulk of the system and that perturb considerably the LSC, to a steady laminar state where a LSC steady coherent structure is markedly dominant and plumes are absent. Based on the analysis of flow patterns at changing the inclination angle, we classified the system behaviour into three regimes: i) turbulent, ii) bursting and iii) laminar.
Despite the major differences observed in the flow organization, the measured variations of $\rm Nu$ are only less than $20\%$ at $\rm Ra=10^8$ and $10\%$ at $\rm Ra=5\times10^8$, while the variations of $\rm Re$ are larger than $85\%$ in all cases. 
More specifically, the Nusselt number shows a non-monotonic two-peak dependency on varying the inclination angle. Such a feature is clearly confirmed both in simulations and in the experiments. The $\rm Re(\beta)$ instead has a a single peak occurring at relatively small angles $\sim 10^o$ followed by a sharp decrease and a near flat behaviour for large inclinations.
In the turbulent regime, it is found that the first peak of $\rm Nu(\beta)/Nu(0)$ is due to the enhancement of the LSC, specifically the enhancement of its component in the direction of the imposed thermal gradient ($y$-component). 
The dependence of $\rm Nu$ and $\rm Re$ on large inclinations  found in our study is very different in comparison to previous studies  conducted in a comparable range of $\rm Ra$ values, in identical containers \cite{Guo2014,Guo2017Evolution} (or containers with similar aspect ratio \cite{Shishkina2016a}) but with rather different $\rm Pr$ values.
In the light of this we advance the hypothesis that the high heat-flux observed in our setup in the laminar regime depends primarily the high-Pr value. In such a condition indeed the temperature diffuses much less than the velocity and it is enslaved to the flow structure and therefore efficiently transported by the laminar LSC; in dimensionless terms the characteristic P\'eclet number of the flow is here always much larger than one, a condition which is not encountered in the previous mentioned studies.\\
\indent 
In summary this study tells that the global heat transfer of thermal convection in a cell with the present features, i.e. $\Gamma =1$, $\rm Pr \gg 1$, $\rm Ra\simeq 10^8 $ and $\rm Ra\simeq5\times10^8$ is robust and is not sensitive to the cessation of turbulence. Furthermore, the laminar heat transfer state is not only equally efficient but also steadier than the turbulent one.  Our results contradict the common wisdom that turbulent thermal convection can be much more efficient in transferring heat than a laminar flow and, in this respect, are in agreement with previous studies at $\Gamma=1$ in different conditions.
The present study also confirms the complex character of the CIC dynamics, which deserves to be explored in its full parameter space in forthcoming studies.
	
	\begin{acknowledgments}
	We thank Yantao Yang and D. Lohse for useful discussions. This work is financially supported by the Natural Science Foundation of China under Grant No.11672156. E.C. acknowledges partial support from the Sino-French (NSFC-CNRS) joint project (No. 11611130099, NSFC China; PRC 2016-2018 LATUMAR, CNRS France).
	\end{acknowledgments}
	
	\nocite{*}
%

\end{document}